\documentclass[preprint,superscriptaddress,amsmath,amssymb,aps,prc]{revtex4-1}
\usepackage{tikz}
\usepackage{graphicx}% Include figure files
\usepackage{dcolumn}% Align table columns on decimal point
\usepackage{csquotes}
\usepackage{braket}
\usepackage{bm}% bold math
\begin{document}

%\preprint{APS/123-QED}

 \title{Measurement of the total neutron scattering cross section ratios of noble gases of natural isotopic composition using a pulsed neutron beam}

\author{Christopher C. Haddock}
%\affiliation{High Energy Accelerator Research Organization KEK 1-1 Oho, Tsukuba, Ibaraki, Japan, 305-0801}
\affiliation{National Institute of Standards and Technology, 100 Bureau Dr., Gaithersburg MD 20899}

\author{Masayuki Hiromoto}
\affiliation{Research Center for Nuclear Physics, Osaka University 10-1 Mihogaoka, Ibaraki, Osaka, 567-0047} 

\author{Katsuya Hirota}%discuss
\affiliation{Nagoya University, Furocho, Chikusa Ward,
Nagoya, Aichi Prefecture 464-0814, Japan}

\author{Takashi Ino}%discuss
\affiliation{High Energy Accelerator Research Organization KEK 1-1 Oho, Tsukuba, Ibaraki, Japan, 305-0801}

\author{Masaaki Kitaguchi}
\affiliation{Nagoya University, Furocho, Chikusa Ward,
Nagoya, Aichi Prefecture 464-0814, Japan}

\author{Kenji Mishima}
\affiliation{High Energy Accelerator Research Organization KEK 1-1 Oho, Tsukuba, Ibaraki, Japan, 305-0801}

\author{Noriko Oi}
\affiliation{Nagoya University, Furocho, Chikusa Ward,
Nagoya, Aichi Prefecture 464-0814, Japan}

\author{Tatsushi Shima}
\affiliation{Research Center for Nuclear Physics, Osaka University 10-1 Mihogaoka, Ibaraki, Osaka, 567-0047} 

\author{Hirohiko M. Shimizu}
\affiliation{Nagoya University, Furocho, Chikusa Ward,
Nagoya, Aichi Prefecture 464-0814, Japan}
\author{W. Michael Snow}
\affiliation{Department of Physics, Indiana University
727 E. Third St., Swain Hall West, Room 117, Bloomington, IN 47405-7105}
\author{Tamaki Yoshioka}
\affiliation{Research Center for Advanced Particle Physics, Kyushu University
744 Motooka, Nishi-ku, Fukuoka, Japan}

\begin{abstract} %%%%%%%%%%%%%%%%%%%%%%%%%% Abstract %%%%%%%%%%%%%%%%%%
Precision measurements of slow neutron cross sections with atoms have several scientific applications. In particular the n-$^{4}$He s-wave scattering length is important to know both for helping to constrain the nuclear three-body interaction and for the proper interpretation of several ongoing slow neutron experiments searching for other types of neutron-atom interactions. We present new measurements of the ratios of the neutron differential scattering cross sections of the noble gases He, Ar, Kr, and Xe, to Ne. All gases used were of natural isotopic-abundance. These measurements were performed using a recently developed neutron scattering apparatus for gas samples located on a pulsed slow neutron beamline which was designed to search for possible exotic neutron-atom interactions and employs both neutron time of flight information and a position\,-\,sensitive neutron detector for scattering event reconstruction. We found agreement with the literature values of scattering cross sections inferred from  Ar/Ne, Kr/Ne and Xe/Ne differential cross section ratios over the $q$ range of $1 - 7$\,nm$^{-1}$. However for the case of He/Ne we find that the cross section inferred  differs by 11.3\% (7.6\,$\sigma$) from previously-reported values inferred from neutron phase shift measurements, but is in reasonable agreement with values from other measurements. The very large discrepancy in the He/Ne ratio calls for a new precision measurement of the n-$^{4}$He scattering length using neutron interferometry. 
\end{abstract}

\maketitle

\section{Introduction} %%%%%%%%%%%%%%%% Introduction %%%%%%%%%%%%%%%%%%
\label{sec:intro}

Precision measurements of slow neutron-atom cross sections have several scientific applications\,\cite{Nico2005, Dubbers2011}. The neutron atom scattering cross section is sensitive to the neutron-nucleus interaction, the neutron-electron interaction, the effects of the neutron electric polarizability from the large electric field experienced by the neutron near the nucleus, and new interactions from possible exotic forces\,\cite{Koester1977}. These different interactions all contain a different dependence on the neutron incident energy and on the momentum transfer to the atom and can therefore be separated experimentally. For light nuclei up to carbon, the neutron-nucleus s-wave scattering lengths which determine the cross section are now of interest to constrain the nuclear few-body force since theory can now calculate the effects of the well-measured NN interaction from first principles\,\cite{Pieper2001, Hagen2010, Tsukiyama2011, Barrett2013, Soma2013}. Both the neutron-electron interaction and the neutron electric polarizability are of interest for the important information they convey about the internal electromagnetic structure of the quarks in the neutron, whose understanding is a major goal for direct calculation from quantum chromodynamics using lattice gauge theory\,\cite{Alexandrou2017, Sufian2017, Abramczyk2017, Shintani2019}. The dependence of the neutron-atom cross section on the incident energy, momentum transfer, and mass of the atom can be used to search for possible exotic interactions of the neutron from new weakly-coupled interactions mediated by exchange bosons with meV-eV masses\,\cite{Nez08}, from short-range modifications to the gravitational interaction from extra dimensions of spacetime\,\cite{Murata2015}, and from certain models for dark matter\,\cite{Fichet2018,Brax:2019koq}. It is therefore always scientifically useful to improve the precision of these measurements. 

Measurements on one of the nuclei presented in this work, $^{4}$He, are of particular scientific interest. Although the values for the two s-wave neutron scattering lengths $b_{\pm}$ corresponding to the scattering amplitudes in the two angular momentum channels $J=I \pm 1/2$ where $I$ is the nuclear spin cannot be calculated for arbitrarily large nuclei at the present time, the nucleon-nucleon (NN) interaction is now measured with enough precision that neutron-nucleus scattering amplitude calculations at low energy in few body nuclei such as H, D, $^{3}$H, $^{3}$He, and $^{4}$He can be compared with experiment to give important information on the poorly-constrained three-body NNN interaction, which is known to be important in nuclei as it is now understood to be responsible for about 10\% of the nuclear binding energy in few nucleon systems. This need has motivated several precision experiments in few body nuclei over the last two decades using neutron pseudomagnetic precession~\cite{Zimmer2002, VanDerBrandt2004} and neutron interferometry~\cite{Black2003, Schoen2003, Huffman2004, Huber2009, Huber2014} to access the spin-dependent and spin-independent components of the s-wave n-A scattering amplitudes. The single n-$^{4}$He scattering amplitude $b_{4}$ from the $I=0$ $^{4}$He nucleus is of particular interest for the interpretation of many slow neutron measurements which search for other types of neutron interactions. The first numerical solution of the five-body Fadeev-Yakubovsky equations relevant for n-$^{4}$He scattering was published in 2018 by Lazauskas and Song~\cite{Lazauskas2018}, who also published a more recent calculation ~\cite{Lazauskas2019} of the parity-odd neutron spin rotation rotary power $d\phi \over dz$ in n- $^{4}$He, which has been sought experimentally in an effort to parametrize the weak interaction in the low energy non-perturbative regime of QCD \cite{Snow2011, Swanson2019}. The result of this recent calculation gives a different result compared to past calculations for the P-odd asymmetry in this observable and also makes a prediction for the n-$^{4}$He scattering length which can be compared to measurement.

The interpretation of a recent experiment using ultracold neutron (UCN) upscattering in $^{4}$He gas as a probe of possible exotic interactions~\cite{Serebrov} also relies on the knowledge of this amplitude. Measurements in progress of the neutron-electron interaction and searches for possible exotic Yukawa interactions of the neutron with atoms using the q-dependence of scattering from noble gas atoms\,\cite{Haddock} will eventually benefit from a high precision measurement of the strong n-$^{4}$He scattering amplitude. 

Unfortunately measurements of $\sigma_{{^4}He}=4\pi b_{4}^{2}$ using different techniques (neutron interferometry~\cite{Kaiser}, neutron transmission~\cite{Rorer}, neutron refraction~\cite{McReynolds}, and UCN upscattering~\cite{Serebrov}) disagree by 10\%, which is unacceptably large for all of the applications referred to above. It is also scientifically embarrassing as for a $I=0$ nucleus like $^{4}$He the technique of neutron interferometry is quite capable of determining the scattering length and therefore the total s-wave scattering cross section with much better than $10^{-3}$ absolute accuracy. The main goal that motivated the analysis of the measurements presented in this paper is to contribute to the resolution of this inconsistency in the experimental data on $^{4}$He. The sensitivity of our measurements, which approach 0.3\% precision, are more than sufficient for this purpose.
 
We have measured the ratio of the differential scattering cross sections of the noble gases He, Ar, Kr, and Xe to Ne, all of natural isotopic-abundance, by performing neutron scattering measurements on the Neutron Optics and Physics (NOP) cold neutron beam line located at the Material Life Science Facility (MLF) at J-PARC. The instrumention for these measurements has been described in detail in~\cite{Haddock}, where it was used to search for deviations from the inverse square law of gravity by studying the momentum transfer ($q$) dependence of neutrons scattered by noble gases using neutron time-of-flight, complementing a similar measurement done earlier at a continuous beam reactor neutron source\,\cite{Kamiya2015}. Gases were chosen for this measurement because the neutron dynamic structure factor $S(q, \omega)$ can be calculated analytically in the ideal gas limit~\cite{Marshall1971}, thereby making it possible to conduct a quantitative analysis of the angular distribution of the scattering and look for deviations from the dominant s-wave contribution from the neutron-nucleus interaction. 

Our instrument was not designed to measure transmitted neutron intensity so we did not conduct absolute measurements of the total scattering cross section. However the relative measurements of the differential cross section that we present are very valuable as several systematic uncertainties cancel in the ratio due to the use of the identical instrumentation and sample environment for the different gas samples. Examples of potential systematic uncertanties which this measurement is relatively insensitive to include, but are not limited to, the absolute pressure and temperature measurements needed to infer the number density of the gas, the absolute knowledge of the thickness of the gas sample container windows, the absolute knowledge of various types of neutron detector backgrounds and electronic offsets, etc. As most of the other measurements conducted in these nuclei were performed in isolation using completely different apparatus and techniques, differences between the differential scattering cross section ratios we present and the values from previous measurements are more likely to indicate the possible presence of uncontrolled systematic uncertainties. Neon of natural isotopic-abundance seems to have the most accurately determined and internally consistent scattering cross section data from previous work and for this reason it was chosen as the sample to normalize all of the cross section measurements from the other gas samples. In addition, to our knowledge this is the first such set of neutron differential cross section ratio measurements on the noble gases conducted using a pulsed neutron source, where one can use neutron time-of-flight information to improve event selection and signal/background separation. Previous work by Krohn et al.~\cite{Krohn73}, which was motivated by an attempt to measure the neutron-electron interaction, also measured scattering cross section ratios from noble gas samples. However this measurement was conducted at a continuous beam reactor neutron source without the benefit of the use of a broad set of neutron energies for additional systematic uncertainty suppression. Our measurement also shares some common approaches with an earlier high-precision measurement of the $n-^{3}$He total cross section in the epithermal energy range from 1 eV to 1 keV~\cite{Keith2004} that was also performed at a pulsed neutron source, but by contrast our measurement was performed using meV neutron energies at nonzero scattering angles rather than eV-keV neutron energies in transmission.

\section{Theoretical Background} %%%%%%%%%%%%%%%% Methodology %%%%%%%%%%%%%%%%%%
\label{sec:method}

The total coherent neutron-atom scattering amplitude for the case of an unpolarized neutron incident upon a fixed, isolated noble gas atom can be written in terms of the momentum transferred from the neutron to the atom during scattering, $q$, as\,\citep{sears}
\begin{equation}
b_\text{c}(q) = b_\text{c}+b_\text{E}(q)+b_\text{M}(q)
\label{eq:btot}
\end{equation}

\noindent where $b_\text{c}$ is the $q$-independent low energy s-wave nuclear scattering amplitude from the strong interaction, $b_{\text{E}}(q)$ describes interactions between the neutrons charge distribution and the atomic electric field, $b_\text{M}(q)$ arises from interactions between the neutrons magnetic dipole moment and the slowly varying electric and magnetic fields of the scattering centers. For the case of diamagnetic atoms (such as the noble gases) with very low incoherent scattering cross sections the contribution from $b_\text{M}(q)$ to the differential cross section is at most $10^{-6}$ times the size of the nuclear contribution, making it negligible compared to our experimental uncertainty. The electric scattering amplitude is written as 
\begin{equation}
b_\text{E}(q) = -b_\text{e} Z f(q)\,,
\label{eq:be}
\end{equation}

\noindent where $b_\text{e} = -1.32(4)\times\,10^{-3}$\,fm\,\citep{KWK,RussiaNE} is the neutron-electron scattering amplitude, $Z$ is the atomic charge number, and $f(q)$ is the atomic form factor which can be computed to sufficient precision using a relativistic Hartree - Fock approximation, whose results are tabulated in the International Tables for Crystallography \cite{ITC}.
\\
\indent The total differential cross section is proportional to the sum of squares of the $q$-dependent coherent and $q$-independent incoherent scattering lengths $b_\text{c}^2(q)$ and $b_\text{i}^2$ respectively. Due to the relatively small value of $b_\text{e}$, the differential cross section be approximated by neglecting terms of $\mathcal{O}(<10^{-3}b_\text{N})$, as $\frac{d\sigma}{d\Omega} \approx b_\text{c}^2+b_\text{i}^2 + 2b_\text{c}b_\text{E}(q)$. In this limit it is clear that the only $q$-dependence comes from the interference between the nuclear and electric scattering amplitudes.  

However since this expression applies only to a fixed and isolated scattering center, it will not accurately describe experimental neutron scattering data which consists of scattering from moving gas atoms which may exchange energy with the neutrons and experience interatomic Van Der Waals - type interactions. A more general expression which accounts for these effects and is sufficiently accurate for our purposes can be written as \cite{Niko}

\begin{equation}
\frac{d\sigma}{d\Omega} = F(q,A,T) \lbrace a_\text{c}^2+a_\text{i}^2 + 2a_\text{c}b_\text{E}(q)\rbrace + (S(q)-1)F(q,2A,T) \lbrace a_\text{c}^2+ 2a_\text{c}b_\text{E}(q)\rbrace
\label{eq:diffRussia} 
\end{equation}
\\
where $a_\text{c}$ and $a_\text{i}$ are the coherent and incoherent \textit{free-atom} nuclear scattering lengths, related to the respective bound values via the ratio of the atomic mass to the neutron mass, $A$, as $a = (\frac{A}{A+1})b$. $F(q,A,T)$ is a kinematical factor which takes into account the thermal motion of the target atoms which are part of an
equilibrium ensemble at temperature $T$. $S(q)$ is the structure factor which describes interference effects arising from atom-atom correlations in the gas coming from Van der Waals interactions and can in principle be calculated using the virial expansion treatment of non-ideal gases. The structure factor in Eq.\,(\ref{eq:diffRussia}) is given for the case of a spherically symmetric potential to first approximation by

\begin{equation}
S(q) = 1 + \frac{4\pi n}{q}\int_{0}^{\infty}
\left(\text{e}^{-U(r)/kT}-1 \right)\sin(qr)\,r\,dr , 
\end{equation}
\\
where $k$ is the Boltzmann constant, $T$ is the temperature, $r$ is the interatomic distance, and $U(r)$ is the interatomic potential. We chose to use the ordinary Lennard-Jones (or \enquote{six-twelve}) potential to describe the interatomic interactions with the parameters given in \cite{lennardjones}. Although there exist several realistic interatomic potentials to model this interaction \cite{Yuri} we found that the difference among them was undetectable in our relatively low and narrow $q$ range for the statistical sensitivity of our measurement. The total cross section can be computed by simply integrating Eq.\,(\ref{eq:diffRussia}) over the entire $4\,\pi$ solid angle. 

\section{Methodology} %%%%%%%%%% Methodology %%%%%%%%%%%%

The technique we use to determine the total scattering cross sections via differential scattering cross section measurements is as follows:

\begin{enumerate}
\item Acquire angular and energy-dependent scattering data for each gas and form the ratios of each spectrum with respect to Ne.
\item Perform a Monte Carlo simulation using the experimentally determined energy spectrum and beam divergence, and literature values of the scattering cross sections as inputs in order to reproduce the spectra obtained in (1).
\item Form a ratio of the experimental and simulated results in (1) and (2) to determine our measured value for the scattering cross section, which is given by an overall shift of the spectra.   
\end{enumerate}

Forming the ratio between two gases replaces the requirement of knowing absolutely the thickness of the gas sample container windows and energy- and scattering angle-dependent efficiencies of our neutron detector, with that of relative knowledge, which greatly surpresses the effect of possible related systematic uncertainties. This means however that our technique is not capable of an absolute scattering cross section measurement and is thus reliant upon the literature for one of the gas species in order to infer cross section values of the remaining gases. Ne currently has the most precise cross section values in the literature (\cite{Rorer}, \cite{Rauch}, \cite{Stehn}), and so it was chosen as the reference gas. 

\subsection{Experimental Setup}

We performed our experiment on a simple scattering apparatus located on the Low-Divergence beam branch of the NOP\, beamline at J-PARC \citep{MLF,NN,design} whose peak energy was measured to be 11.7 meV at the time of measurement. The beam power during data acquisition was approximately 410 kW. The essential components consist of a gas cell, an evacuated scattering chamber (also called the \enquote{vacuum chamber}) and a\,$^3$He position sensitive detector (PSD). The entrance and exit windows of the gas cell were made of $0.1$\,mm-thick aluminum windows. When filled with gas, the cell was pressurized to approximately 1.85\,atm. Between the measurement of each gas species the cell was evacuated and scattering data was collected for the empty cell condition. This allowed for monitoring of the stability of the apparatus between runs which allowed us to ensure that components in the beamline had not shifted between fills.

\begin{figure}[!ht]
\centering
\includegraphics[width = 0.75\textwidth]{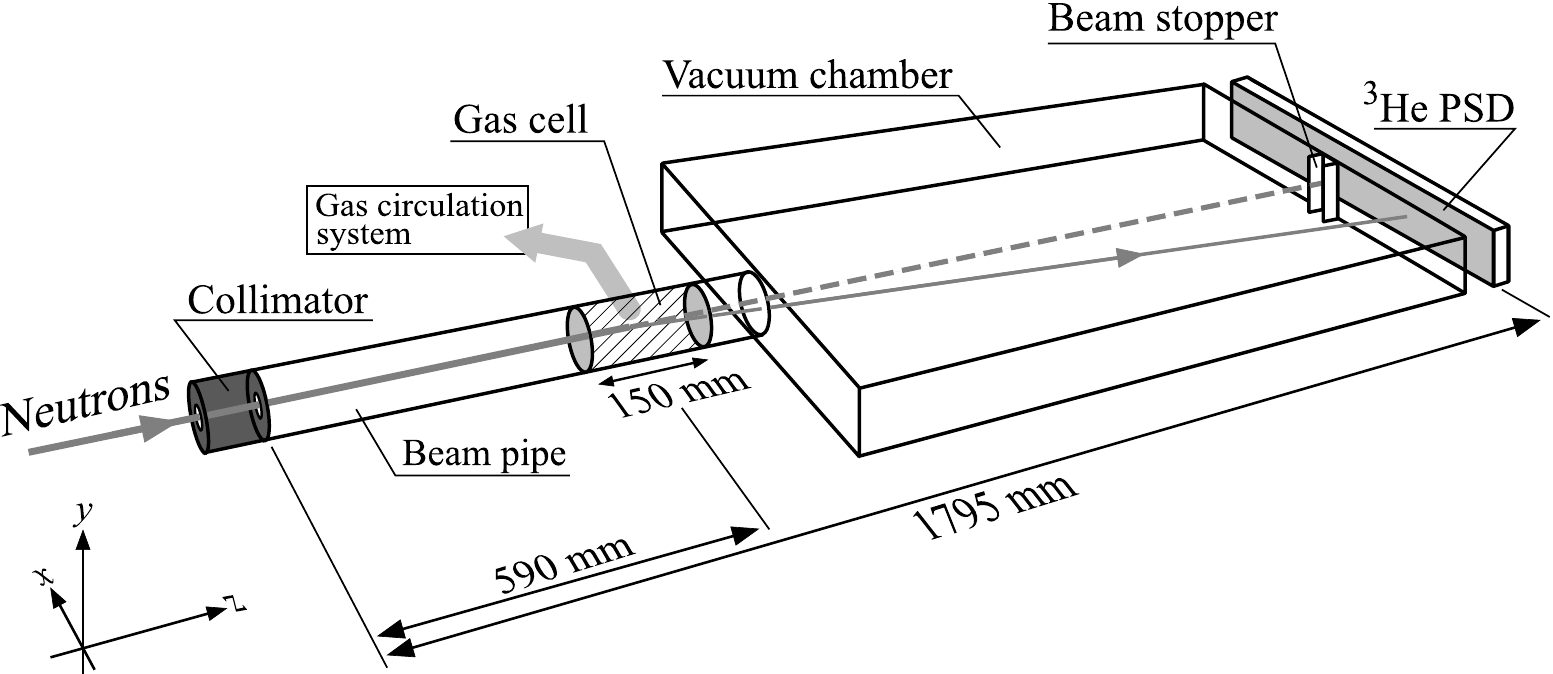}
\caption{Layout of our experiment as mounted on BL05 at the MLF
facility at J-PARC. Image taken from \cite{Haddock}. }
\label{fig:apparatus}
\end{figure}

The experimental layout is shown in Figure \ref{fig:apparatus}. A more detailed description of the functionality of the apparatus can be found in\,\cite{Haddock} where the $q$-dependence of low energy neutron-noble gas scattering was studied in efforts to search for possible deviations from the inverse square law of gravity. The region of $q$ used in the present measurement which is dictated by the neutron energy spectrum and the experimental geometry is $1 - 7$\,nm$^{-1}$.

Before taking gas scattering data we mapped the intensity distribution in the incident neutron beam using a $1 \times 1$ mm$^2$ collimator formed from two sets of neutron absorbing B$ _4$C plates. By recording the data as a function of slit position in the x-y plane, a two-dimensional intensity and time-of-flight distribution was obtained. These distributions are used as input for the Monte Carlo simulation. 
%This data was used to produce the plots
%shown in Figures \ref{fig:vspec} and \ref{fig:sscan}.

% \input{400_simulation}
\subsection{Simulation}
\label{sec:simulation}
The experiment was simulated using the Monte Carlo method implemented within the ROOT analysis framework. Neutrons are generated in a loop and assigned an energy and position chosen from the two-dimensional scans of neutron intensity of the beam described in the previous section. Neutrons are propagated to and from scattering centers using standard kinematic relations. The interaction point is determined by the total interaction cross section used as input. If an interaction occurs, it is determined to be scattering or absorbing according to their relative probabilities. If a neutron scatters from a gas atom near the upstream edge of the cell it is less likely to reach the detector due to the reduced solid angle, which may have a small effect on the measured neutron scattering angular dependence, distinct from the neutron-atomic interactions. It is therefore important that we select input cross sections that are close to the expected values so that the angular dependence is sufficiently reproduced. We found that using existing literature values were sufficient for this purpose.     

Once scattered, the neutron angular distribution is determined from Eq.(\ref{eq:diffRussia}) and is propagated to the detector according to the experimental geometry. The ROOT framework then allows us to store and view the final phase space coordinates using appropriate time-of-flight and angular cuts so that the results could be directly compared with experimental data.% Values for the scattering lengths used to reproduce the q spectra from Eq.(\ref{eq:diffRussia}) were taken from \cite{sears92} for
%He, Ne, and Ar while values for Xe and Kr were taken from
%\cite{Krohn73}. 

\section{Results} %%%%%%%%%% Experimental Results %%%%%%%%%%%%

The neutron count rate was measured separately for each gas species and for the empty cell. Empty-cell data is subtracted from that of the gas-filled cell and subsequently normalized by the number density (as measured via independent pressure and temperature measurements) and the proton beam power at MLF during data aquisition. The results are displayed in Table \ref{tab-1}. Corrections to the ideal gas law were made when computing the number density using the second virial coefficients given in \cite{He2vir} for He, and \cite{gas2vir} for the other gases.

To determine the value of the cross section we first consider the general expressions for the experimental and simulated count rates neglecting the possibility of multiple scattering events (See Sec. \ref{sec:su}),

\begin{align}
  R_{\text{EXP}}(q) &= \epsilon_{\text{EXP}}(q)\frac{d\sigma(q)}{d\Omega}\label{eq:Rexp}\\           
  R_{\text{SIM}}(q) &= \epsilon_{\text{SIM}}(q)\frac{d\sigma(q)}{d\Omega}\label{eq:Rsim}
\end{align}                                                                                      
                                                                                                 
\noindent where $\epsilon(q)$ describes any q-dependence which arises from the cell-detector geometry (e.g. scattering from an extended object vs. point) and detector efficiency. If there were no q-dependence due to the detector efficiency, the simulation and the experiment would reveal the same q spectra for a given species differing only by a constant factor. This however was not the case as the detector efficiency was found to increase slightly as a function of q, due to an increase in mean free path of the neutron in the $^3$He detector with an increase in scattering angle, a gas species\,-\,indepdendent effect. Since the exact form of this function was not known for our detector, to determine the scattering cross section we formed the following ratio between two gases, one of which was always Ne, causing the respective $\epsilon(q)$ terms to cancel:                                              
% form the ratio of the measured scattered neutron count rate for gas species $R$ and Ne, $R_{\text{Ne}}$. We then form the same ratio, but with simulated data. Using Eq.\,\ref{eq:sigma} below we extract the measured scattering cross section making use of the fact that the count rate $R_{\text{SIM}}$ is directly proportional to $\sigma_{\text{SIM}}$:\\

\begin{equation}
  \sigma_{\text{MEAS}}=
  \left(\sum\limits_i\frac{R(q_i)}{R_{\text{Ne}}(q_i)}\right)_{\text{EXP}}
  \left(\sum\limits_i\frac{R(q_i)}{R_{\text{Ne}}(q_i)}\right)^{-1}_{\text{SIM}}  
  \times \sigma_{\text{SIM}}\,. 
\label{eq:sigma}
\end{equation}\\

\noindent where the sums are carried out over q values (120 total in our measurement) in the range from $q = 1 - 7$\,nm$^{-1}$. 

\iffalse
\begin{equation}
\sigma_{\text{MEAS}}= \left(\frac{R}{R_{\text{Ne}}}\right)_{\text{EXP}}  \left(\frac{{R}}{R_{\text{Ne}}}\right)^{-1}_{\text{SIM}} \times \sigma_{\text{SIM}}\,. 
\label{eq:sigma}
\end{equation}\\
\fi
  
Except for the case of Ne, we do not propagate errors in the literature associated with those input values used in the simulation as they only serve as scaling factors. However since the reference value for Ne remains in the expression used to determine the measured cross sections, the errors are propagated to the listed simulated count rate ratios, and therefore to the measured scattering cross sections reported later in Sec. \ref{subsec:csv}.

\begin{table}[!h]
\begin{tabular}{l l l l l l}
\hline
Gas & $R[\text{s}^{-1}]$ & $R/R_{\text{Ne}}$ & $(R/R_{\text{Ne}})_{\text{SIM}}$ & $\sigma_{f,\text{SIM}}$\cite{sears92,Krohn73} \\\hline
He & 7.192(16) & 0.5958(18) & 0.6721(17) & 0.857 \\
Ne&  12.070(35) &      $-$ &      $-$ &  2.383(6) \\
Ar & 2.855(18) & 0.2365(15)& 0.2366(6)& 0.648\\
Kr & 29.041(50) &2.406(6) &2.443(6) & 7.61\\
Xe & 16.110(40) & 1.3347(39) & 1.3422(32) & 4.30\\\hline
\end{tabular}
\caption{The measured normalized count rate at 0.5 MW of beam power and gas cell pressurized to 2 atm for $q$ in 1-7\,nm$^{-1}$ is given in the second column. Rates relative to the measured Ne data are given in the third column.  Simulated rates with respect to Ne rates obtained using the value $\sigma_b^{Ne} = 2.628(6)$ \cite{sears92}, converted to the free-atom value in the simulation, are given in the fourth column. The input free-atom cross sections used in the simulation are given in the fifth column. Uncertainties in the experimental data originate from statistics only. Uncertainties in the simulated data arising from the propagation of the literature uncertainty for Ne dominate the statistical uncertainty in the simulated data.}
\label{tab-1} 
\end{table}

The count rate alone is not enough to infer the cross sections of the gases as there are contributions from the thermal motion of the gas atoms, the atomic electric field, and the interatomic potential experienced by the gas atoms. We compared the ratio of the count rate of a gas species with respect to Ne (Table \ref{tab-1}, column 3) to the respective ratio of the simulated values (Table \ref{tab-1}, column 4) using the values in column 5 as input free-atom scattering cross sections.\\
\indent The values used as inputs for the scattering cross sections and scattering lengths were taken from \cite{sears92} for He, Ne, and Ar, and from \cite{Krohn73} for Xe and Kr. It would appear that the measured He count ratio is significantly lower than predicted by simulation as compared to the other gas species, but consistent with the older literature results from neutron transmission which will be discussed later.

\begin{figure}[!h]
\centering
\includegraphics[width=0.75\textwidth]{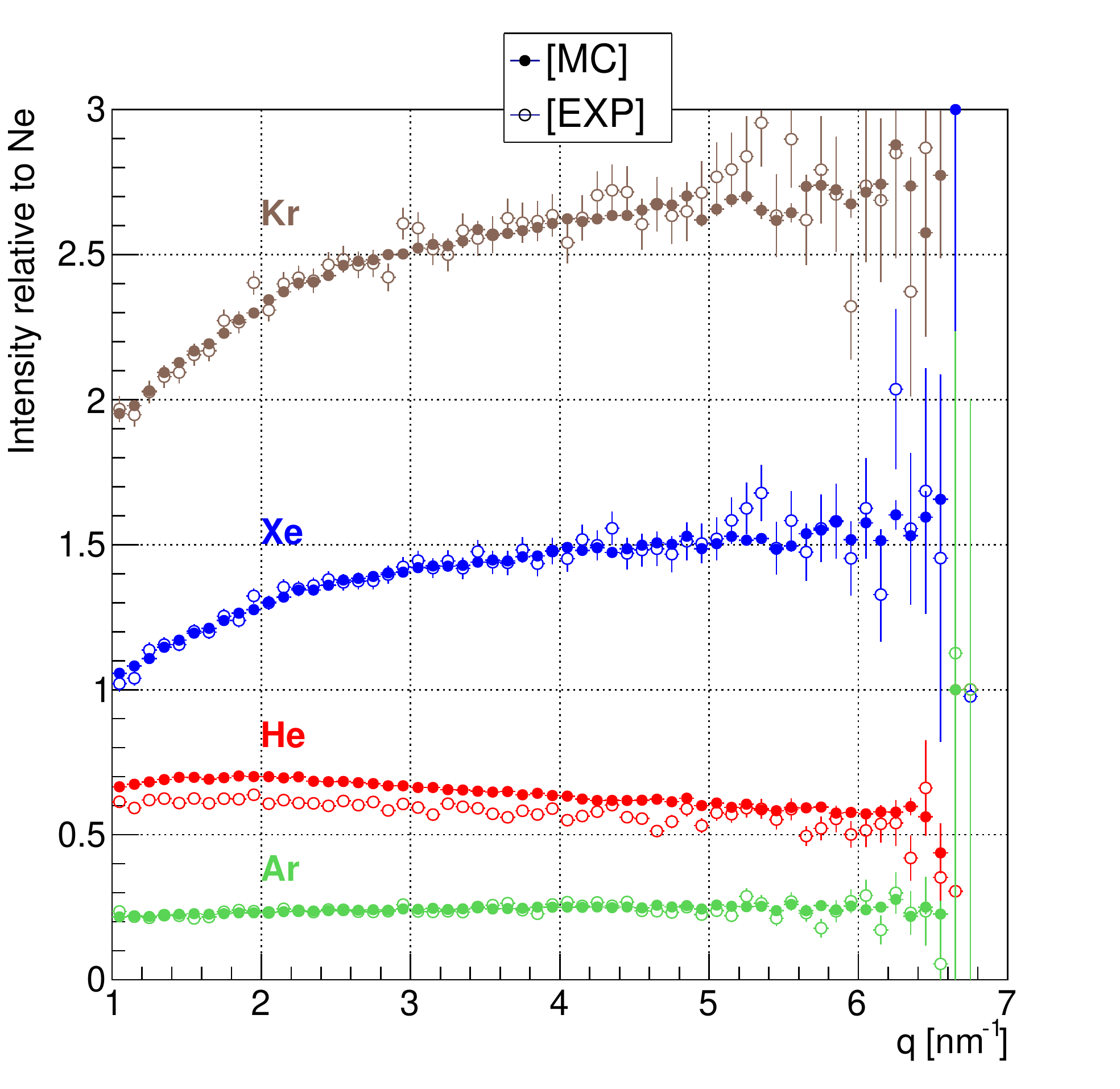}
\caption{Ratios of experimental and simulated data with respect to
Neon. }
\label{fig:rawrat}
\end{figure}

\subsection{Systematic Uncertanties}
\label{sec:su}

\subsubsection{Pressure and Temperature Stability}
The pressure and temperature of the gas sample was measured continuously throughout each run using Mensor CPG2400 digital pressure gauge, a Pfeiffer PKR251 ion gauge, and a PT100 platinum resistance thermometer with accuracies of 300 Pa (0.15\%) at 2\,atm, and 60 mK (0.02\%) at 300\,K, respectively.
\\
\indent The average values of pressure and temperature for each run were used to determine the number density, which was in turn used to normalize the data to give a measured intensity per atom in the gas target. The pressure and temperature data was verified to either decrease or increase monotonically throughout each run so that the average value used to infer the target density is correct to first order. Higher order effects arising from the change the temperature and pressure dependence of the scattering function used in the simulation are less than $10^{-5}$ in magnitude and thus negligible.

\subsubsection{Pressure dependence of cell geometry}
A possible source of unwanted systematic scattering uncertainties may arise from the fact that, when pressurized, the thin Al windows of the gas cell become slightly distorted relative to an unpressurized, evacuated cell. Since empty-cell data was subtracted from gas-filled data in our analysis it was necessary to quantify the size of this effect. The distortion of the Al windows may give rise to a change in the measured $q$ spectra and/or total transmission of the neutron beam, both due to the slight change in thickness of the warped window.
\\
\indent To estimate the change in thickness of the Al window, we first measured the deflection of the center of the window after pressurization to $\sim 2$ atm and found it to be no more than 3\,mm. We then use the computed ratio of the unpressurized and pressurized surface areas as a reduction factor to the thickness of the window, assuming that the window volume remains constant during pressurization. The reduction in thickness in the region of the window seen by the neutron beam was found to be $0.03\%$. When considering only the change in beam attenuation from nuclear scattering, a negligibly small $\sim 10^{-5}$ effect in the measured cross section is found when forming the ratio of vacuum subtracted measurements with respect to Ne.
\\
\indent Another effect which may arise due to the change in thickness of the Al windows is that of inelastic single-phonon scattering. The size of inelastic single-phonon scattering in the differential cross section is approximated from measurement for the case of Al at room temperature in \cite{Barker}. A roughly constant effect on the intensity of $I(q) = 5.2 \times 10^{-4}$/cm/sr was found, which when converted to a microscopic differential cross section is $8.7 \times 10^{-3}$ bn/sr. Given our solid angle acceptance of $\Omega = 0.034$ the effect is $2.96 \times 10^{-4}$ and as it is applied only to the change in window thickness computed above, makes a negligible contribution to our uncertainty in our cross section measurement. 

\subsubsection{Multiple Scattering}

If a neutron interacts with more than one gas atom before reaching the detector, Eq.\,\ref{eq:Rexp} and Eq.\,\ref{eq:Rsim} are not exactly correct and our method of extracting the cross section via Eq.\,\ref{eq:sigma} may fail. Fortunately, the probability of multiple scattering is very small given the gas pressures used and the geometry of our setup. For example, the probability of a single scattering even from an ideal gas of cross section $\sigma$ and number density $n$ in a cylinder of length $L$ is given by $p_{\text{1}} = 1 - e^{-\rho\sigma L}$. If we consider the gas with the largest measured scattering cross section Kr, with $\sigma_f = 7.61$, the scattering probability in a typical run is $p_{\text{1}} \approx 4\times 10^{-3}$, or $0.4\%$. This value can provide a rough estimate for the relative likelihood of multiple to single scattering. In that case we see that this effect may be on the order of our statistical precision and must be corrected for via the computation of correction factors for each gas.

To compute the correction factors, the original simulation code was adjusted to account for multiple scattering events under the assumption of isotropic scattering to significantly reduce computation time for practicality. This is justified since the scattering function is isotropic to $<1\%$ for all gases in our region measured region of $q$. The correction factors were found to fall in the range of $[2-11]\times 10^{-4}$, causing a shift in the measured cross section values by $<0.1\%$ in our measurement. These corrections are included in our reported values in Table \ref{tab-2}.

%We can then very roughly assume the probability of a second scattering event to be on the order of $p_{\text{1}}^2 \sim 10^{-5}$ which, although small, is approaching the region of statistical sensitivity of our measurement.   

%Although our simulation accounted for scattering events followed by absorption events, due to the high likelihood of absorption for Xe and Kr, it did not take into account multiple scattering events. 

%To determine the size of the effect of specifically multiple scattering events on the measured 

%To determine an upper bound on the size of the effect of specifically multiple scattering events we consider the gas with the largest measured scattering cross section Kr, with $\sigma_f = 7.61$. The probability of a single scattering event for a neutron entering parallel to the target cell (no beam divergence) is given by $p_{\text{1}} = 1 - e^{-\rho\sigma L_0}$, where $\rho$ is the number density of the gas, $\sigma_s$ is the scattering cross section and $L_0$ is the length of the cell. For 2 atm of Kr in a 15\,cm-long cell, the product $\rho\sigma L_0$ is much less than 1 so we can write $p_{\text{1}} \approx \rho\sigma L_0 = 3.8\times 10^{-3}$.    

%Computing the probability $p_{\text{ms}}$ of multiple scattering gives %$p_{\text{ms}} \leq (1-e^{-\rho\,\sigma_fL})^2$. Substituting $\rho = 5\times10^{19}\text{cm}^{-3} $ for the number density at 2 atm and $L = 10$\,cm for the cell thickness gives  $p_{\text{ms}} = 3.85\times10^{-3}$.

\subsubsection{Uncertainty in Simulation}

The statistical uncertainty in the Monte Carlo simulation is computed in exactly the same manner as in the experiment, where the number of simulated scattered neutrons reaching the detection region, $N$, follow a Poisson distribution and thus contribute $1/\sqrt{N}$ to the counting error. The number $N$ for the simulated gases was at least an order of magnitude larger than the number of experimentally detected neutrons so that the contribution from simulation statistical error was relatively small.

In our analysis we form the ratio between vacuum-subtracted experimental data for the gases while the simulated data accounts for scattering only from the gas (it does not include background scattering from the cell windows). Because each gas has a different transmission probability, the vacuum data must be scaled appropriately for each gas before subtraction so that scattering from the beam stop and from the downstream aluminum cell window and vacuum chamber flange is completely removed. This was done by scaling the vacuum run energy spectra using the attenuation factor $e^{-\rho\,\sigma_TL}$ where $\rho$ is the number density, $\sigma_T$ is the total interaction cross section (scattering plus absorption), and L is the cell thickness. Uncertainties in values for the total cross section are small but not entirely negligible for the cases of Xe and Kr whose absorption cross section uncertainties are relatively large. For Xe, the absorption cross section $\sigma_{abs}$ is 25.1(1.0) bn \cite{Krohn73} which translates to an uncertainty of $0.023\%$ in our determination of the scattering cross section. Likewise for Kr,  $\sigma_{abs}$ is 25.0(8) which translates to an uncertainty in $\sigma_{s}$ of $0.013\%$. These uncertainties were propagated to the final cross section values, however uncertainty in the transmission values for the remaining gases was less than  $10^{-5}$ and thus negligible.   

The purpose of forming the ratio between two simulated gas scattering results is to determine the relative angular dependence of the scattering between two gases so that the only difference between the ratio of two gases in a simulation compared to the experiment is a constant multiple whose magnitude indicates the deviation of the experimental cross section from the theoretical input. Error in the $q$-dependence of the simulation resulting from uncertainties in the values of the neutron electron scattering legnth, the coherent and incoherent scattering lengths, and the Lennard-Jones parameters used to compute the interatomic pair potential were determined by varying the input parameters over the published uncertainty and comparing the resulting $q$ spectra. The distributions changed at the level of $10^{-5}$ and thus contribute negligibly in the determination of the cross section value at the level of 0.1\%.
\subsection{Cross Section Values} %%%%%%%%%%%%%%% Results and Discussion %%%%%%%%%%%%%
\label{subsec:csv}

\iffalse
%without multiple scattering corrections
\begin{table}[!h]
%\caption{Cross Section Measurements}
\begin{tabular}{l|llllll}
  \hline
Gas & He & Ar & Kr & Xe & \\\hline
A & 3.997 & 37.998 & 83.04 & 129.99 \\ \hline
%$\sigma_{\text{b,Ref}}$[bn] & & & & \\ 
$\sigma_{\text{f}}$[bn] &  0.7601(30)& 0.648(4) & 7.495(26) & 4.276(16) \\
$\sigma_{\text{b}}$[bn] & 1.188(5)  & 0.683(5) & 7.677(26) & 4.342(17) \\
\end{tabular}
  \caption{Values of the total scattering cross sections determined through differential cross section measurements made relative to the Ne value $\sigma_{\text{b}}^{\text{Ne}} = 2.628(6)$\,bn\,\cite{sears92}. The relation between free and bound atom cross section is determined using $\sigma_{\text{f}}=(\frac{A}{A+1})^2\,\sigma_{\text{b}}$, where A is the ratio of the atomic mass to the neutron mass, computed in \cite{Krohn73}, and given in the second row of this table.}
\label{tab-2} 
\end{table}
\fi

%with multiple scattering corrections
\begin{table}[!h]
%\caption{Cross Section Measurements}
\begin{tabular}{l|llllll}
  \hline
Gas & He & Ar & Kr & Xe & \\\hline
A & 3.997 & 37.998 & 83.04 & 129.99 \\ \hline
%$\sigma_{\text{b,Ref}}$[bn] & & & & \\ 
$\sigma_{\text{f}}$[bn] &  0.7599(30)& 0.648(4) & 7.503(26) & 4.278(16) \\
$\sigma_{\text{b}}$[bn] & 1.188(5)  & 0.683(5) & 7.685(26) & 4.344(17) \\
\end{tabular}
  \caption{Values of the total scattering cross sections determined through differential cross section measurements made relative to the Ne value $\sigma_{\text{b}}^{\text{Ne}} = 2.628(6)$\,bn\,\cite{sears92}. The relation between free and bound atom cross section is determined using $\sigma_{\text{f}}=(\frac{A}{A+1})^2\,\sigma_{\text{b}}$, where A is the ratio of the atomic mass to the neutron mass, computed in \cite{Krohn73}, and given in the second row of this table.}
\label{tab-2} 
\end{table}

The values of the total free-atom scattering cross section are obtained from our data by forming the ratio of the $q$ spectra of the experimental and simulated data integrated over all measured $q$ values. This ratio is then multiplied by the scattering cross section used in the simulation for the non-Ne gas to obtain the measured cross section of that gas. The value of the scattering cross section for Ne is assumed to be 2.628(6) bn (taken from \cite{sears92}) in our analysis.

\begin{figure}[!ht]
\centering
\includegraphics[width = 0.65\textwidth]{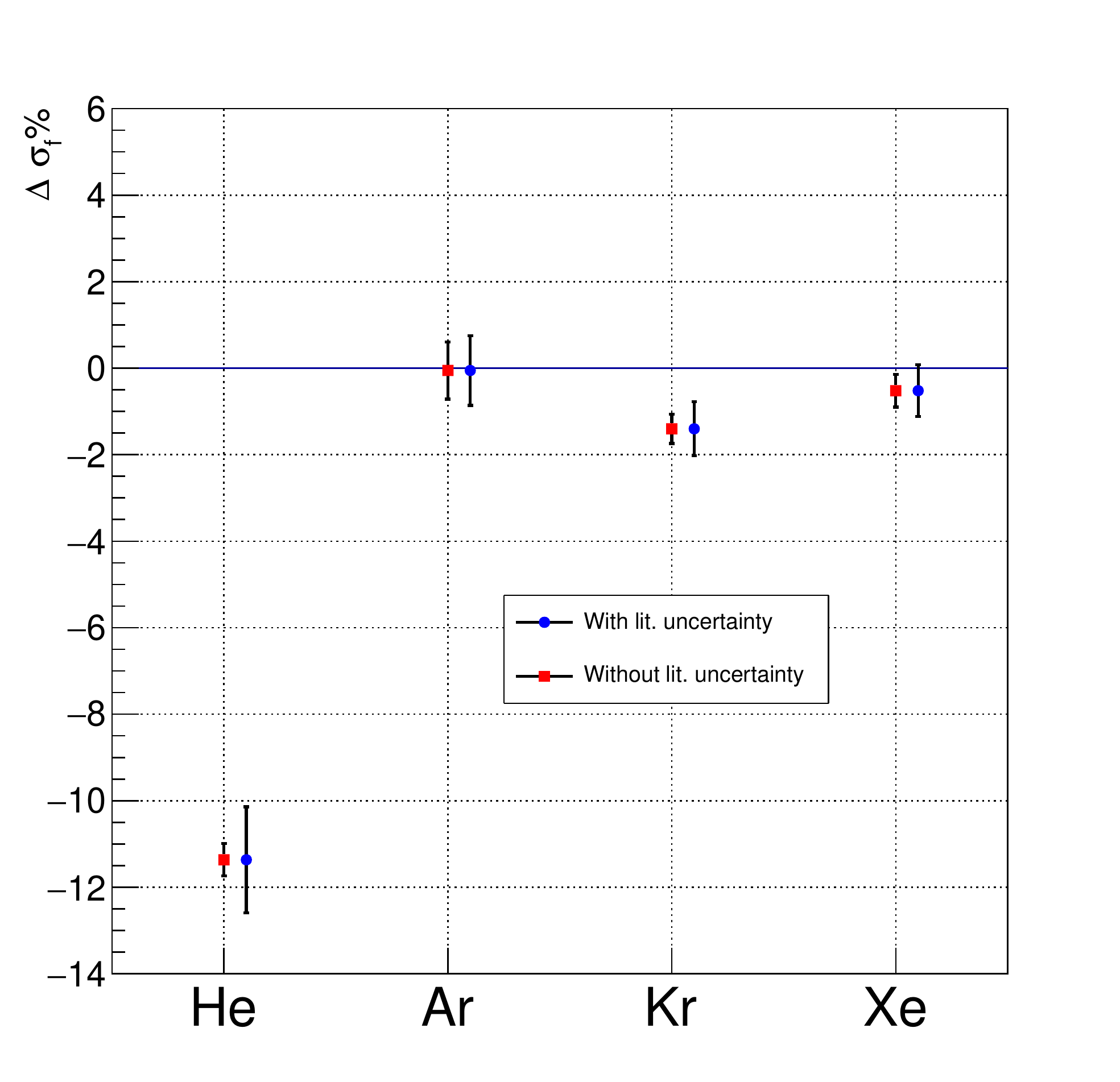}
\caption{The percent difference between our measured values and those in the literature for the free-atom scattering cross sections, with and without propagation of uncertainties from the literature. Our values were compared with $\sigma_{\text{f}}^{\text{He}} = 0.857(10),\sigma_{\text{f}}^{\text{Ar}} = 0.683(4),$ using values from \cite{sears92}, and  $\sigma_{\text{f}}^{\text{Kr}} = 7.61(4),\sigma_{\text{f}}^{\text{Xe}} = 4.30(2),$ using values from \cite{Krohn73}. We chose to compare Kr and Xe to \cite{Krohn73} rather than the values in \cite{sears92} due to the lower level of uncertainty for Kr and the availability of the incoherent scattering contribution for Xe in \cite{sears92}. }
\label{fig:CWL}
\end{figure}

Our results for the ratio of the differential cross sections of Ar, Kr, and Xe with respect to Ne are consistent with existing literature values, both from neutron transmission and neutron interferometry measurements \cite{Rorer, sears92, Mug, Krohn73}. For the case of Xe and Kr our values are the most precise. However the case for He is very different from the results from neutron interferometry but are in agreement with the other previous measurements using transmission and UCN upscattering \cite{Rorer,Serebrov2}. The results are summarized in Table \ref{tab-2} and are compared with free atom cross sections in the literature in Figure \ref{fig:CWL}.\\  
\indent Our measured bound scattering cross section, $\sigma_{\text{b}}^{\text{He}} = 1.188(5)$ is in disagreement with the value $\sigma_{\text{b}}^{\text{He}} = 1.34(2)$ in \cite{sears92}, a discrepancy of 7.6 standard deviations. The bound n-$^{4}$He scattering length corresponding to our result in Table \ref{tab-2} is $b_c^{\text{He}} =3.075(6)$\,fm. This differs from the value $b_c^{\text{He}} = 3.26(3)$\,fm from the neutron interferometry measurement in \cite{Kaiser} which is listed in a compilation \cite{sears92} which is often quoted in the literature, and is the same reference from which we are taking our Ne reference value from.\\
\indent It is worth noting that if one wanted to know the results of our cross section measurements if a different value for the Ne cross section was used in the normalization, a very good approximation ($<1\%$) can be made by mutiplying the value of our measured cross section to the ratio of the Ne cross section in \cite{sears92} to the alternative reference value. For example, in the Atlas of Neutron Resonances \cite{Mug} the free atom Ne cross section is given as $\sigma_{\text{f}}^{\text{Ne}} = 2.415(10)$. Using this value instead as a normalization results in a value of $b_c^{\text{He}} \approx 3.09$\,fm, still many standard deviations from the interferometric measurements of $b_c^{\text{He}}$, but in agreement with the transmission and UCN upscattering values mentioned above.\\
\indent Given our result and its consistency with other n-$^{4}$He measurements, we strongly suspect that there is an unaccounted-for systematic uncertainty that somehow crept into the neutron interferometry result in n-$^{4}$He. It is therefore very timely that a new neutron interferometry measurement of the coherent scattering length of n-$^{4}$He, conducted mainly with the motivation to help constrain the three-nucleon interaction, has been recently carried out at the Neutron Interferometry and Optics Facility (NIOFa) at the NIST Center for Neutron Research, described in a recent Ph.D thesis \cite{Rob}, however the results are not published at the time of this writing.\\
\indent A change in the accepted value of $\sigma(^4 $He) is also relevant for the results of a recent analysis which searched for possible exotic interactions of the neutron using upscattering of ultracold neutrons from helium gas~\cite{Serebrov}. If our results are confirmed by subsequent measurements then the conclusions of this analysis may need to be reexamined.

\section{Conclusion} %%%%%%%%%%%%%%%%%%%%%%% Conclusion %%%%%%%%%%%%%%%%%%%%%%%%%%

We performed neutron differential scattering cross section measurements of the noble gases He, Ar, Kr, and Xe, relative to Ne on BL05 at the MLF facility at J-PARC. The values of the total scattering cross sections inferred from our measurements are consistent with the literature values of Ar, Kr, and Xe, where for Kr and Xe our measurements are the most precise. However our data suggests a value for the total scattering cross section for n-$^{4}$He which is in disagreement with existing neutron interferometry data, but consistent with older data based on transmission and reflectometry. A new neutron interferometry measurement of n-$^{4}$He from NIST has been carried out however the results are not yet published. The value from this new measurement will be important for nuclear few body theory and for the proper interpretation of a number of ongoing and planned measurements of neutron electroweak interactions with atoms and also for different neutron searches for possible exotic interactions. 

The other noble gas which deserves higher precision neutron scattering length measurements in the near future is xenon. Better data on this nucleus would be helpful for the interpretation of the ongoing J-PARC noble gas scattering measurements already mentioned above. Natural isotopic abundance xenon has several stable isotopes, but only two of these stable isotopes, $^{129}$Xe and $^{131}$Xe, have nonzero nuclear spin. Therefore three measurements would be sufficient for a complete characterization of the coherent and incoherent scattering amplitudes of the natural isotopic abundance mixture of xenon gas which will be used in the J-PARC gas scattering measurements. One of these measurements can be neutron interferometry. The same apparatus used for n-$^{4}$He can simply be filled with xenon to conduct this measurement. The other two measurements would need to be conducted on polarized samples of $^{129}$Xe and $^{131}$Xe. Both of these nuclei can be polarized using spin-exchange optical pumping in amounts sufficient that one could perform a polarized neutron pseudomagnetic precession measurement to determine the difference $a_{+}-a_{-}$ of the s-wave scattering lengths in these nuclei \cite{walkerSEOP,GoodsonNMR,StupicNMR}.

\section*{Acknowledgments} %%%%%%%%%%%%%%%%%%% Acknowledgments %%%%%%%%%%%%%%%%%%%%%%%%

This work was supported by MEXT KAKENHI grant number JP19GS0210 and JSPS KAKENHI grant number JP25800152. We wish to thank the help given by Setsuo Sato for detector and software operation.
Work at the facility of J-PARC was performed under an S-type project of KEK (Proposal No. 2014S03) and user programs (Proposal No. 2016B0212, 2016A0078, and 2015A0239). C. Haddock acknowledges support from the Japan Society for the Promotion of Science. C. Haddock and W. M. Snow acknowledge support from NSF grant PHY-1614545 and from the Indiana University Center for Spacetime Symmetries.

\bibliographystyle{plain}

\end{document}